\documentstyle[preprint,aps]{revtex}

\input psfig

\tightenlines

\begin{document}

\draft

\title{\bf STUDIES OF QUANTUM SPIN LADDERS\\
AT $T=0$ AND AT HIGH TEMPERATURES\\
BY SERIES EXPANSIONS}

\author{J. Oitmaa$^1$\cite{byline1}, Rajiv R.P Singh$^2$\cite{byline2},
 and Zheng Weihong$^1$\cite{byline3}} 
\address{${}^1$School of Physics,                                              
The University of New South Wales,                                   
Sydney, NSW 2052, Australia.\\                      
${}^2$Department of Physics,                                              
University of California,                                   
Davis, CA 95616, USA.}                      

\date{Feb. 7, 1996}

\maketitle 

\begin{abstract}
We have carried out extensive series studies, 
at $T=0$ and at high temperatures,  of 2-chain and 3-chain
spin-half ladder systems with 
antiferromagnetic intrachain and both antiferromagnetic and 
ferromagnetic interchain couplings.
Our results confirm the existence of a gap in the
2-chain Heisenberg ladders for all 
non-zero values of the interchain couplings.
Complete dispersion relations for the spin-wave excitations
are computed. For 3-chain systems, our results are consistent
with a gapless spectrum. We also calculate
the uniform magnetic
susceptibility and specific heat
as a function of temperature. We find that as $T\to 0$, for
the 2-chain system the uniform susceptibility goes rapidly 
to zero, whereas for the $3$-chain system it
approaches a finite value. These results are compared in detail
with previous studies of finite systems.
\end{abstract}                                                              


\narrowtext
\section{INTRODUCTION}
The magnetic properties of low dimensional systems have been  the subject
of intense theoretical and experimental research in recent years.
It is by now well established that one-dimensional Heisenberg antiferromagnets
with integer spin have a gap in the excitation spectrum, whereas 
those with half-integer spin  have gapless excitations.
The former have a finite correlation length, while
for the latter it is infinite with the spin-spin
correlation function decaying to zero as a power law.
In 2-dimensions, the unfrustrated square-lattice Heisenberg model
has long range N\'{e}el order in the ground state. It has gapless
Goldstone modes as expected.
In recent years much interest has focussed on systems
with intermediate dimensionality and on questions of crossovers
between $d=1$ and $d=2$. One approach to this problem has been
to study a two-dimensional system where the coupling for the
spins separated along the $x$-axis is different from those
separated along the $y$-axis
\cite{affleck}. It has been suggested that an alternative
way to explore this issue is through
the Heisenberg spin ladders consisting of a finite number of
chains coupled together, with a coupling $J_{{}_{/\!/}}$ along 
the chains and $J_{\perp}$ between them.  These systems have been the subject
of considerable recent theoretical and experimental interest.

Experimentally,
2-chain $S={1\over 2} $ ladders are realized in vanadyl pyrophosphate
 $(VO)_2P_2O_7$\cite{ecc94} and in the strontium cuprate
$SrCu_2O_3$\cite{tak92}, whereas 
3-chain $S={1\over 2} $ ladders are realized  in the 
strontium cuprate
$SrCu_3O_5$\cite{tak92}.

Theoretically, a number of striking predictions have been 
made for such systems. These have
been recently reviewed by Dagotto and Rice\cite{dag95}.
Barnes {\it et al.}\cite{bar93,bar94} were the first to
carry out extensive Monte Carlo
studies of the excitation spectrum and the magnetic susceptibility
for 2-chain ladders with antiferromagnetic interchain coupling.
White\cite{whi94} {\it et al.} and Hida\cite{hid95} have used 
the density matrix renormalization method to study the
spin gap. Watanabe\cite{wat94} has applied the numerical
diagonalization method to finite systems of 2-chain ladders with ferromagnetic
interchain coupling. Azzouz\cite{azz94} {\it et al.} developed
a mean-field theory
and used the density-matrix renormalization group 
method to study 2-chain ladders.
Gopalan, Rice and Sirgrist\cite{gop94} presented  a 
variational wavefunction 
for the ground state of the 2-leg ladder. Rojo\cite{roj95} has 
shown that  spin ladders
with odd number of legs  have gapless excitations. 
Troyer {\it et al.} \cite{tro94} have used improved versions of the quantum
transfer-matrix algorithm to study the temperature 
dependence of the susceptibility,
specific heat, correlation length etc of 2-chain ladders.
Finite-size scaling was used by Hatano\cite{hat95} for 
multi-leg ladders, and recently Frischmuth
{\it et al.}\cite{fri96} and Sandvik {\it et al.}\cite{san95} have
applied Quantum Monte Carlo simulation to compute the temperature dependence
of uniform susceptibility and internal energy for spin ladders with up to
6 legs.
One clear result from all these studies is that
ladders with an even number of legs have an 
energy gap, short range correlations and a ``spin liquid" ground
state. On the other hand, ladders with odd number of legs have 
gapless excitations, quasi long range order, and a power-law falloff of
spin-spin correlations, similar to single chains. Experiments also
confirm these features.

We have carried out extensive series studies of 2-chain and 3-chain
ladder systems with both antiferromagnetic and ferromagnetic interchain coupling
$J_{\perp}$, via Ising expansions and dimer expansions at $T=0$, and
also by high temperature series expansions. 
Our results confirm the existence of a 
gap in the 2-chain system and delineate the phase diagram in
the parameter space of Ising anisotropy and the parameter ratio
$J_\perp/J_{{}_{/\!/}}$.
The complete spin-wave excitation spectra are computed. 
For the 3-chain system
we are unable to exclude the possibility of a small gap, although our results are 
consistent with a gapless spectrum. In addition, we develop a 
high-temperature series
expansion for the uniform magnetic susceptibility and the specific heat for 2-chain and 3-chain
systems with $J_{{}_{/\!/}}=J_{\perp}$; the susceptibility of 2-chain ladders is as expected for a system with a spin-gap
while that of 3-chain ladders 
appear to remain finite in the zero-temperature limit suggesting 
an absence of a spin-gap.
We compare our results in detail with previous calculations.

\section{Series Expansions}
The Hamiltonian of a spin ladder with $n_l$ legs  is given by,
\begin{equation}
H= J_{{}_{/\!/}} \sum_{i,l=1}^{l=n_l} {\bf S}_{l,i} \cdot {\bf S}_{l,i+1}
 + J_{\perp} \sum_{i,l=1}^{l=n_l-1} {\bf S}_{l,i}\cdot {\bf S}_{l+1,i}  \label{H}
\end{equation}
where ${\bf S}_{l,i}$ denotes the S=1/2 spin at the $i$th site of the $l$th
chain.
$J_{{}_{/\!/}}$ is the interaction between nearest neighbor spins 
along the chain and $J_{\perp}$ is the interactions between 
nearest neighbor spins along the rungs. We denote the ratio of
couplings as $y$, that is, $y\equiv J_{\perp}/J_{{}_{/\!/}} $.
In the present paper the 
intrachain coupling is taken to be antiferromagnetic
(that is, $J_{{}_{/\!/}} > 0$) whereas the interchain coupling
$J_\perp$ can be either antiferromagnetic or ferromagnetic. This
includes the values of interest in the experimental systems
discussed earlier where $J_{\perp}\sim J_{{}_{/\!/}}$. 
Without loss of generality,
we can set $J_{{}_{/\!/}}=1$ hereafter.

We have carried out three different expansions for the system, The first is
 the expansion about the Ising limit at zero
temperature for both two and three-chain ladders.
We have computed  the
ground state properties as well as the spin-wave 
excitation spectra by this expansion.
The second is the dimer expansion, again at $T=0$. This expansion
can be done  for the 
2-chain system with antiferromagnetic interchain coupling only. 
The third is the high-temperature series
expansion for the uniform susceptibility of
the 2-chain and 3-chain ladders with $y=1$.

\subsection{Ising expansions}
To perform an expansion about the Ising limit for this system, we
introduce an anisotropy parameter $x$, and write the Hamiltonian
in Eq.(\ref{H}) as:
\begin{equation}
H = H_0 + x V  \label{Hising}
\end{equation}
where 
\begin{eqnarray}
H_0 &= & \sum_{i,l=1}^{l=n_l} S_{l,i}^z S_{l,i+1}^z +
  y \sum_{i,l=1}^{l=n_l-1} S_{l,i}^z S_{l+1,i}^z  \nonumber \\
V &= &  \sum_{i,l=1}^{l=n_l} ( S_{l,i}^x S_{l,i+1}^x + S_{l,i}^y S_{l,i+1}^y )  + 
y \sum_{i,l=1}^{l=n_l-1} ( S_{l,i}^x S_{l+1,i}^x + S_{l,i}^x S_{l+1,i}^x ) 
\end{eqnarray}
The limits $x=0$ and $x=1$ correspond to the Ising model and
the Heisenberg model respectively.
The operator $H_0$ is taken as the unperturbed
Hamiltonian, with the unperturbed ground state being the 
usual N\'{e}el state
for  antiferromagnetic interchain coupling, 
and a fully ordered collinear state for 
 ferromagnetic interchain coupling.
The operator $V$ is treated as  a perturbation.
It flips a pair of spins on neighbouring sites. 
The Ising expansion method 
has been previously reviewed in several  
articles\cite{he90,gel90}, and will not be repeated here.
The calculations involved a list of 9184 linked clusters of up to
16 sites for the 2-chain ladder, and 14082 linked clusters of up to
12 sites for the 3-chain ladder, together with their lattice constants
and embedding constants.

Series have been calculated for 
the ground state energy per site $E_0/N$,
the staggered magnetization $M$ for $y>0$ (or collinear magnetization $M$
for $y<0$), the parallel staggered/colinear susceptibility $\chi_{{}_{/\!/}}$,
and the uniform perpendicular susceptibility $\chi$ per site for several 
ratio of couplings $y=\pm 0.1, \pm 0.25, \pm 0.5, \pm 0.75, \pm 1, \pm 1.5, \pm 2, 
\pm 4, \pm 8$ up to order $x^{16}$ for 2-chain ladders, and $x^{12}$ for 
3-chain ladders (the
series for uniform perpendicular susceptibility $\chi$ is one order less
in each case).
The resulting series for $y=\pm 1$ for the 2-chain and 3-chain systems are listed in
Tables I and II, the series for other value of $y$ are available on request. 

To analyze these series, we first performed a standard Dlog Pad\'{e}
analysis of the magnetization $M$ and parallel 
susceptibility $\chi_{{}_{/\!/}}$ series.
We found that for 2-chain ladders, the series lead to a simple power-law singularity at 
$x<1$:
\begin{equation}
M \sim (1-x/x_c)^{\beta} \quad \chi_{{}_{/\!/}} 
\sim (1-x/x_c )^{-\gamma}
\end{equation}
with the indices $\beta$ and $\gamma$ close to $1/8$ and $7/4$
respectively. This transition at $x< 1$, with criticality in
the universality class of the $2D$-Ising model, is  strong evidence
that in the Heisenberg limit the system has a disordered ground state
with a spin-gap, as is the case for the spin-one chain \cite{s1}.
In contrast for
the 3-chain ladders, the series analysis showed poor convergence
and suggested a singularity at $x\geq 1$.
This implies that for the 3-chain ladders, the system is analogous to
the spin-half chain, with gapless spectra and power-law correlations
in the Heisenberg limit.

Fig. 1 shows the phase boundary for 2-chain ladders as function
of $y$. It is interesting to note the different behaviour 
for $y>0$ and for $y<0$:
for antiferromagnetically coupled 2-chain ladders ($y>0$), 
$x_c$ decreases as $y$ increases, and
in the limit of $y\to \infty$, $x_c$ will approach 0. But for
ferromagnetically coupled 2-chain ladders ($y<0$),  $x_c$ first
decreases as the absolute value of $y$
increase from zero, but then the trend reverses and
it, once again, approaches 1 as $y\to -\infty$. 
To understand this behavior, we can 
map the system for large negative $y$ 
to a spin-one chain with on-site ( single-ion)
anisotropy:
\begin{equation}
H = {1\over 2} \sum_i [ {\bf S}_{i,tot} \cdot {\bf S}_{i+1, tot} + y (1-x) (S_{i,tot}^z)^2 ]
\end{equation}
where ${\bf S}_{i,tot}$  denotes the S=1 spin at the $i$th site 
of the chain. We can get the asymptotic behaviour of the phase boundary
in the limit of  $y\to -\infty$ by studying the following spin-one
chain with on-site anisotropy:
\begin{equation}
H= {1\over 2} \sum_i [  S_{i,tot}^z S_{i+1, tot}^z - c (S_{i,tot}^z)^2 +
  x' ( S_{i,tot}^x S_{i+1, tot}^x + S_{i,tot}^y S_{i+1, tot}^y ) ]
\end{equation}
For this model,
we have carried out series expansions in $x'$ to order $x'^{14}$
 (i.e. 14 sites) for staggered magnetization $M$ 
for several different value of $c$: $c=0.275, 0.29, 0.3, 0.325$, 
and performed a standard Dlog Pad\'{e}
analysis to find the critical value  $c'$ which gives
the singularity of $M$ at $x'=1$. We get $c'=0.29(1)$. Hence, the 
asymptotic behaviour of the phase boundary in the limit 
$y\to -\infty$ is given by
\begin{equation}
x_c = 1 + 0.29/y
\end{equation}
which is also shown in Fig. 1 as a bold line near $y/(1+|y|) = -1$.

Fig. 2 gives the results of the ground-state energy per site $E_0/N$
versus $y$ for both 2-chain and 3-chain ladders at the Heisenberg
point $x=1$. Our results for the ground state energy agree
extremely well with
the recent quantum Monte Carlo simulation\cite{fri96}.
In Fig. 3, we present the uniform perpendicular susceptibility
at $T=0$.

We also performed the Ising expansion for the triplet spin-wave excitation
spectrum of 2-chain and 3-chain
ladders using Gelfand's method \cite{gelfand}. 
To overcome a possible singularity at $x<1$ 
in the 2-chain ladders, and to get a
better convergent series in the Heisenberg limit, 
we add the following staggered field term
to the Hamiltonian in
Eq. (\ref{Hising}):
\begin{equation}
\Delta H = t (1-x) \sum_i (-1)^i S_i^z
\end{equation}
$\Delta H$ vanishes at $x=1$. We adjust the coefficient $t$ to 
get the most smooth terms in the series, 
with a typical value being $t=2$.
We computed the Ising expansion for the triplet spin-wave excitation spectrum
$\epsilon (k)$ up to order $x^{15}$ for 2-chain ladders, and up to order $x^{11}$ 
for 3-chain ladders. These series are too long to be listed here, 
but are available  on request.

These series have
 been analyzed by using  integrated first-order inhomogeneous
 differential approximants\cite{gut}.
For the 2-chain ladder, there are 2 bands of excitations,
Fig. 4 shows the dispersion
$\epsilon (k)$, with $k_y=\pi$,
for antiferromagnetic interchain coupling. 
The other band with $k_y =0 $ is related to this by $\epsilon (k_x, 0) = 
\epsilon(\pi - k_x , \pi)$. This is simply due to the staggered
field, which doubles the spectrum.
As a comparison, the dispersion relation of a single chain (that is the case of $y=0$)
 is also shown. It can be seen from the graph that in the limit $y \to 0$, the
dispersion relation has a simple cosine function with a period of $2 \pi$, and in the limit
$y\to \infty$, the
dispersion relation also has a simple cosine form
with a period of $4 \pi$, and a gap  in the spectrum.
Barnes and Riera\cite{bar94} argued that the
dispersion, for all $y$, can be fitted by combining these two functions 
into the following form:
\begin{equation}
\epsilon (k_x, \pi)^2 = \epsilon (0, \pi)^2 \cos^2 (k_x/2) + \epsilon (\pi , \pi)^2 
\sin^2 (k_x/2) + c_0 \sin^2 (k_x)
\end{equation}
For $y=1$, the Ising expansions give an energy gap of
$\epsilon (\pi, \pi) = 0.44(7)$. A more precise estimate
is obtained by the 
dimer expansions, which give $\epsilon (\pi, \pi) =0.504(7) $.
We will discuss the dimer expansions later.

For ferromagnetic interchain coupling, the two bands of spectra  are independent, but
 each band is a simple cosine function with a gap at the minimum and 
symmetric about $k_x=\pi/2$, as shown in  Figs. 5-6.
As noted in Figs. 4, it is clear that the spin-gap decreases
smoothly as $y$ is reduced, and vanishes 
at $y=0$. These results agree well with previous calculations\cite{bar94}.

For the 3-chain system, there are three bands. In the 
Ising limit, 2 bands have initial excitations
located in the side rows, and the third band has it
in the middle row. Figs. 7-12 show the
spectrum of the three bands for ferromagnetic and antiferromagnetic interchain
couplings. From these graphs, we can see that all of the dispersion relations have
a simple  cosine function (except for middle row band with 
large $y$) with a minumum located at $k_x=0$ (or $k_x=\pi$ by symmetry),
where two of these three bands have a definite gap, the third
(the $k_y=0$ band) is consistent with a gapless spectra. 
The estimate for the gap in the third band for all $y$ values
is $0.2(3)$ (except for the case of $y=0$ where we got 0.08(10)).
Hence, we cannot  exclude the possibility of a small gap here.
But our results are
consistent with a gapless spectrum, and given our earlier results
on the phase boundary with Ising anisotropy, 
we believe the spectra are gapless.

\subsection{Dimer expansions}
For 2-chain ladders, with antiferromagnetic coupling between 
the chains, there is an alternative $T=0$ expansion
that can be developed. In the limit that the exchange coupling along the rungs $J_{\perp}$
is much larger than the coupling $J_{{}_{/\!/}}$ along the chains, that is $y\gg 1$,
  the rungs interact only weakly
with each other, and the dominant configuration in the ground state is the product
state with the spin on each rung forming a spin singlet, so the Hamiltonian in Eq. (\ref{H})
can be rewritten as,
\begin{equation}
H/J_{\perp} =  H_0 + (1/y) V  \label{Hdimmer}
\end{equation}
where
\begin{eqnarray}
H_0 &=& \sum_{i,l=1}^{l=n_l-1} {\bf S}_{l,i} \cdot {\bf S}_{l+1,i} \nonumber \\  && \\
V &=& \sum_{i,l=1}^{l=n_l} {\bf S}_{l,i} \cdot {\bf S}_{l,i+1}  \nonumber 
\end{eqnarray}
We can treat the operator $H_0$ as the unperturbed
Hamiltonian. The eigenstates of a single pair of spins, or
dimers, consists of one singlet state with total $S=0$ and
eigen energy $E_s = -3/4$:
\begin{equation} 
| \Psi \rangle_s = {1\over \sqrt{2} } (\vert \uparrow \downarrow \rangle - \vert \downarrow \uparrow \rangle )
\end{equation} 
and three triplet states with total $S=1$ and
eigenenergy $E_t = 1/4$:
\begin{equation} 
| \Psi \rangle_t =[ {1\over \sqrt{2} } 
(\vert \uparrow \downarrow \rangle + \vert \downarrow \uparrow \rangle ), ~~
\vert \uparrow \uparrow \rangle,  ~~  \vert \downarrow \downarrow \rangle ]
\end{equation} 
The operator $V$ is treated as  a perturbation.
It can cause excitations on a pair of neighbouring dimers. 
Details of the dimer expansions 
and the matrix elements of $V$ are given in Ref. {\onlinecite{gel90}}, 
and will not be repeated here.

We have carried out the dimer expansion for the ground state energy
to order $(1/y)^9$ and for the lowest lying triplet excitations
to order $(1/y)^8$. 
The series for the ground state energy per site $E_0/N$ is:
\begin{eqnarray}
E_0/N = && J_{\perp} [ -3/8 - 3/(16 y^2) -3/(32 y^3) + 3/(256 y^4) + 45/(512 y^5) + 159/(2048 y^6)
\nonumber \\
&& - 879/(32768 y^7) - 4527/(32768 y^8) - 248391/(2097152 y^9) ]
\end{eqnarray}
and the series for the excitation spectrum are listed in Table III.
Again, we use the integrated first-order inhomogeneous
 differential approximants\cite{gut} to extrapolate the series. For
$y>1 $, the dimer expansions give much better results than the Ising expansions. 
For $y \sim 1$ also the dimer expansions appear to
converge better.
The overall spectra determined from the combined study of dimer 
and Ising expansions are shown in Fig. 4.

\subsection{High temperature series expansions}
We now turn to the thermodynamic properties of the ladder system
at finite temperatures. We have developed
high-temperature series expansions for the uniform magnetic susceptibility
$\chi (T)$ 
and the specific heat $C (T)$,
for 2-chain  and 3-chain system with
$J_{\perp} = J_{{}_{/\!/}} $, 
\begin{eqnarray}
 \chi (T) &=& {\beta \over N} \sum_i \sum_j {{\rm Tr}
 S_i^z S_j^z e^{-\beta H} \over  {\rm Tr} e^{-\beta H } } \nonumber  \\  &&  \\
 C (T) &=& {\partial U \over \partial T} \nonumber
\end{eqnarray}
where $N$ is the number of sites, and $\beta=1/(k_BT)$, and the internal
energy $U$ is defined by
\begin{equation}
U =  { {\rm Tr} H e^{-\beta H} \over  {\rm Tr} e^{-\beta H } }
\end{equation}
The series were computed to order $\beta^{14}$. The number of
contributing graphs, with upto $14$ bonds was $4545$ for the
$2$-chain ladders and $5580$ for the $3$-chain ladders.
The series are listed in Table IV. We use  
integrated first-order inhomogeneous
 differential approximants\cite{gut} to extrapolate the series. 
The resulting estimates are
shown in Fig. 13-14. For the susceptibility,
as a comparison, the recent Quantum Monte Carlo (QMC) results of 
Frischmuth et al
\cite{fri96} and the results from our T=0 Ising expansion for 3-chain are also shown.
It can be seen that our results agree very well with the QMC results except 
for the 3-chain system at very low temperatures. Given the recent
findings that for the spin-half chain, the $T=0$ value is
reached from finite temperatures with infinite slope \cite{affleck2}, one might
expect the $T\to 0$ behavior for these 3-chain systems to be equally 
complex, making it very difficult to explore numerically.
For the specific heat, our results showed
good convergence upto the peak, but poor convergence
below it. The results for 2-chain ladder are
consistent with recent 
quantum transfer-matrix 
calculations by Troyer\cite{tro94} {\it et al}.

\section{CONCLUSIONS}

We have studied the $2$ and $3$ chain Heisenberg-Ising ladders
by a variety of different series expansions. Our results confirm
the existence of a gap in the excitation spectrum of $2$-chain
systems, with either ferromagnetic or antiferromagnetic interchain
interactions. For $3$-chain systems, our results are consistent 
with a gapless spectrum. The $T=0$ phase diagram as well as the
temperature dependence of the uniform susceptibility and the
specific heat are also calculated. Overall, our results agree
very well with previous numerical studies of these systems.

\acknowledgments
This work forms part of a research project supported by a grant 
from the Australian Research Council. R. R. P. S. is supported
in part by the National Science Foundation through grant number
DMR-9318537 and would like to thank the University of
New South Wales for hospitality and the Gordon Godfrey
Foundation for support, while the work was being done.
We would also like to thank Dr. Troyer for providing us with 
the Monte Carlo data for comparison.


\begin{figure}[htb]
\vspace{9pt}
\makebox[80mm][l]{{\hskip 1.3pc} \psfig{file=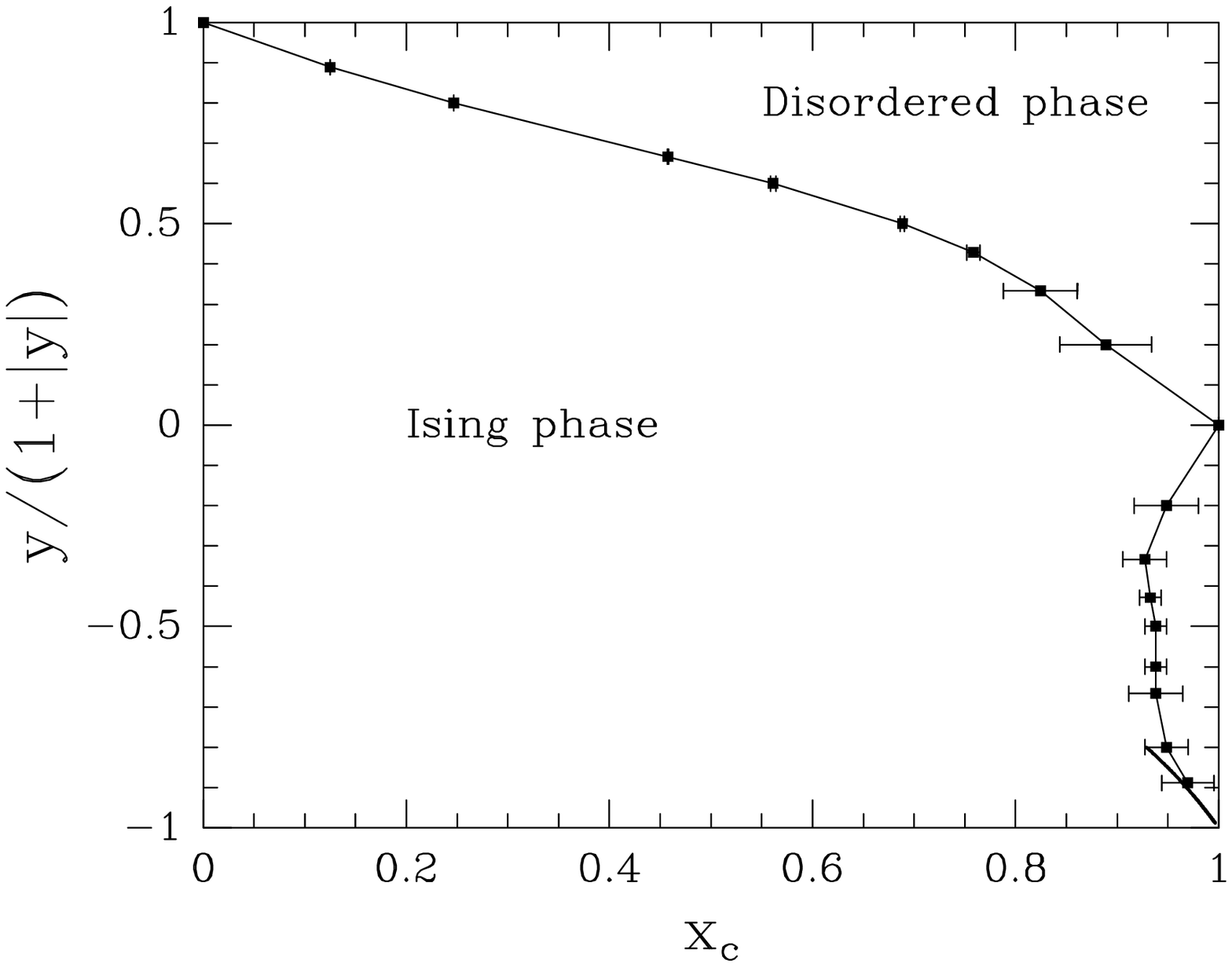,width=80mm}}
\caption{The phase boundary for 2-chain ladder.   
The asymptotic behaviour as $y\to -\infty$ predicted by a spin-one single chain system
 with one-site anisotropy is also shown by the bold line.}
\label{fig:fig1}
\end{figure}

\begin{figure}[htb]
\vspace{9pt}
\makebox[80mm][l]{{\hskip 1.3pc} \psfig{file=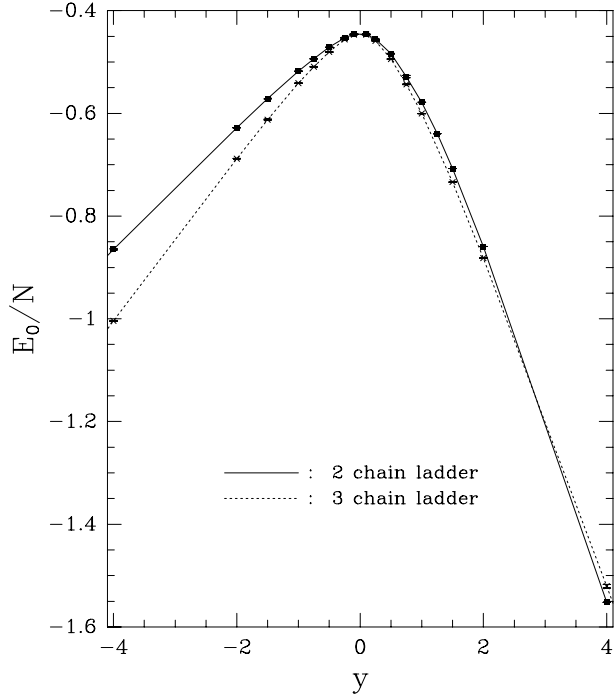,width=95mm}}
\caption{The ground-state energy per site $E_0/N$ as function of $y$ for
2-chain and 3-chain ladders. The error bars are much smaller than
the symbols.
}
\label{fig:fig2}
\end{figure}

\begin{figure}[htb]
\vspace{9pt}
\makebox[80mm][l]{{\hskip 1.3pc} \psfig{file=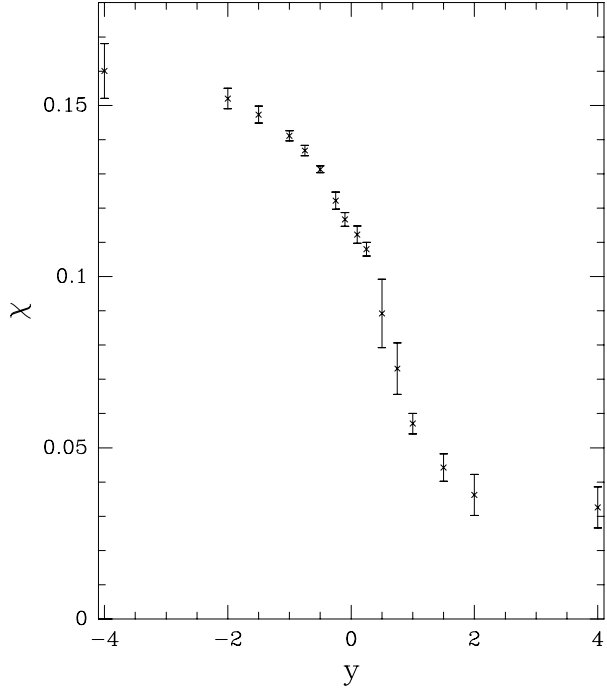,width=95mm}}
\caption{The uniform susceptibility $\chi$ at $T=0$ 
as function of $y$ for the 3-chain ladder.
}
\label{fig:fig3}
\end{figure}

\begin{figure}[htb]
\vspace{9pt}
\makebox[80mm][l]{{\hskip 1.3pc} \psfig{file=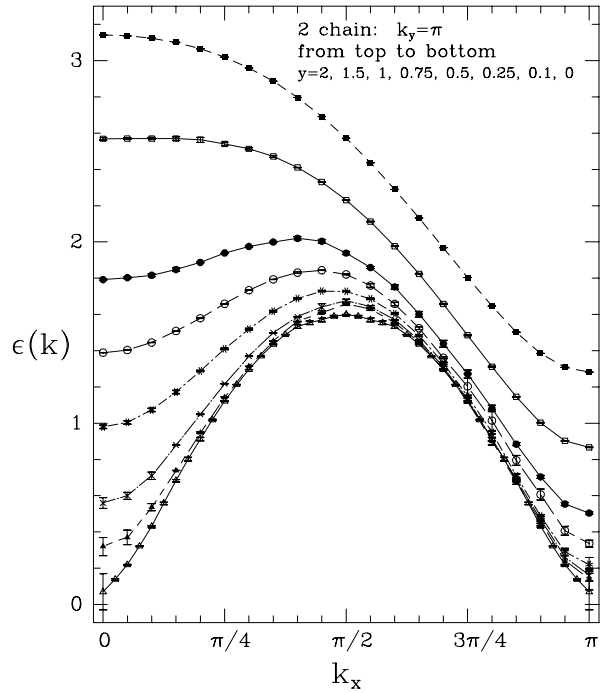,width=95mm}}
\caption{The dispersions of the spin-triplet excitated states of the 2-chain ladder 
with antiferromagnetic interchain coupling $y=$2, 1.5, 1, 0.75, 0.5, 0.25, 0.1, 0 (shown in
the figure from  the top to 
the bottom respectively), for $k_y=\pi$.
}
\label{fig:fig4}
\end{figure}

\begin{figure}[htb]
\vspace{9pt}
\makebox[80mm][l]{{\hskip 1.3pc} \psfig{file=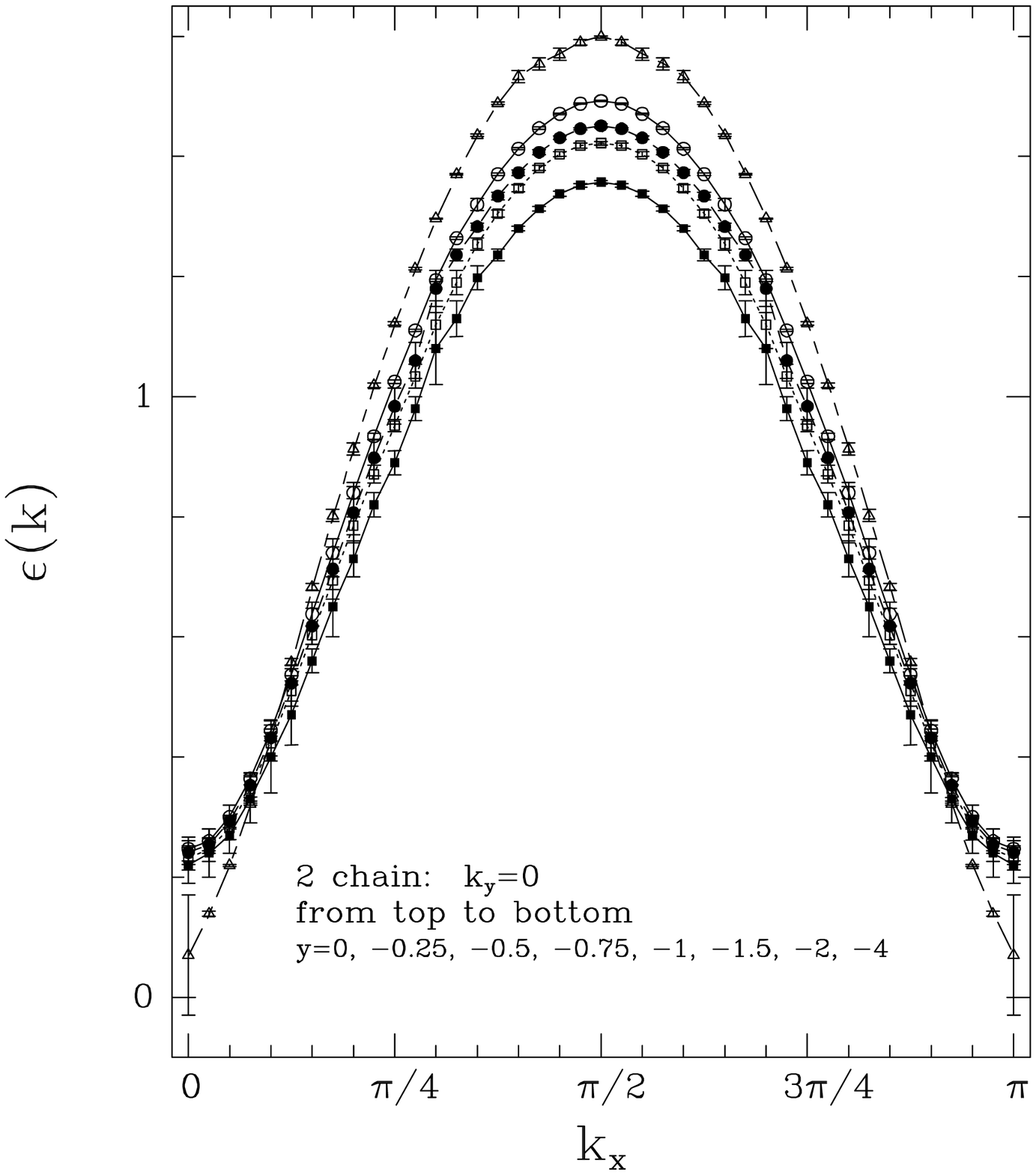,width=95mm}}
\caption{The dispersions of the spin-triplet excitated states of the 2-chain ladder 
with ferromagnetic interchain coupling $y=$0, -0.25, -0.5, -0.75, -1, -1.5, -2, -4
, for $k_y=0$.
}
\label{fig:fig5}
\end{figure}

\begin{figure}[htb]
\vspace{9pt}
\makebox[80mm][l]{{\hskip 1.3pc} \psfig{file=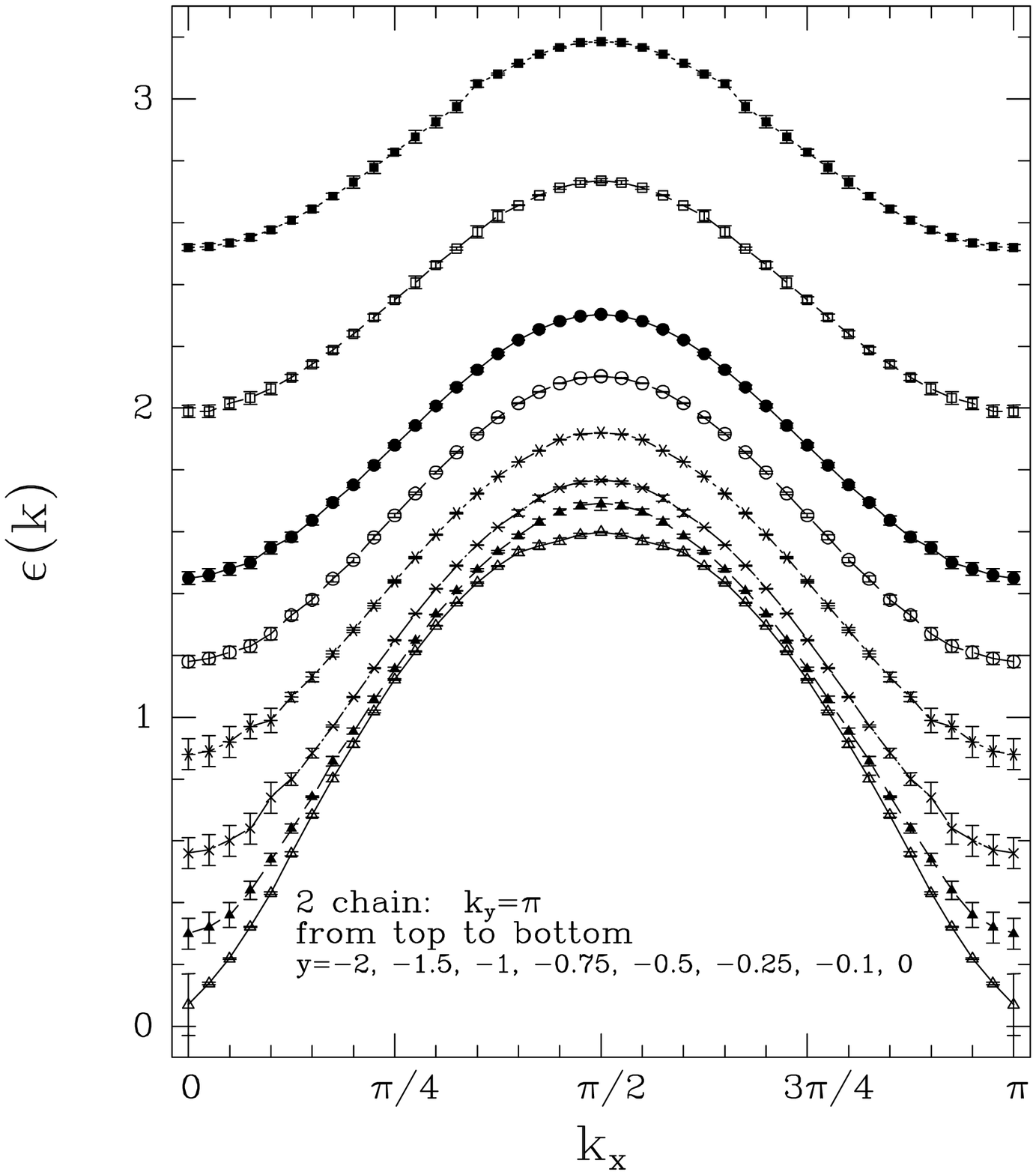,width=95mm}}
\caption{The dispersions of the spin-triplet excitated states of the 2-chain ladder
with ferromagnetic interchain coupling 
$y=$-2, -1.5, -1, -0.75, -0.5, -0.25, -0.1, 0, for $k_y=\pi$.
}
\label{fig:fig6}
\end{figure}

\begin{figure}[htb]
\vspace{9pt}
\makebox[80mm][l]{{\hskip 1.3pc} \psfig{file=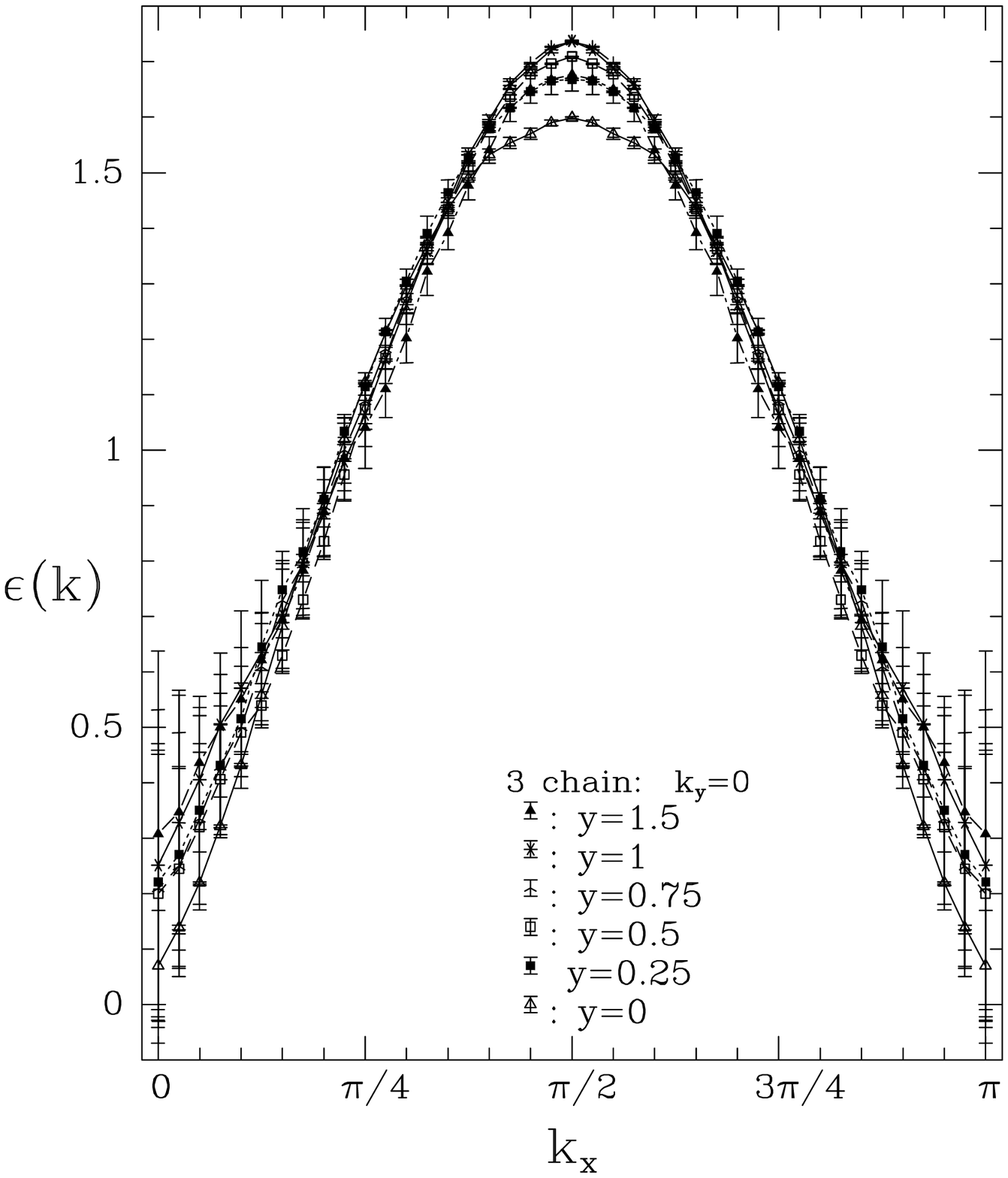,width=95mm}}
\caption{The dispersions of the spin-triplet excitated states of the 3-chain ladder
with antiferromagnetic interchain coupling $y=$1.5, 1, 0.75, 0.5, 0.25, 0, for $k_y=0$.
}
\label{fig:fig7}
\end{figure}

\begin{figure}[htb]
\vspace{9pt}
\makebox[80mm][l]{{\hskip 1.3pc} \psfig{file=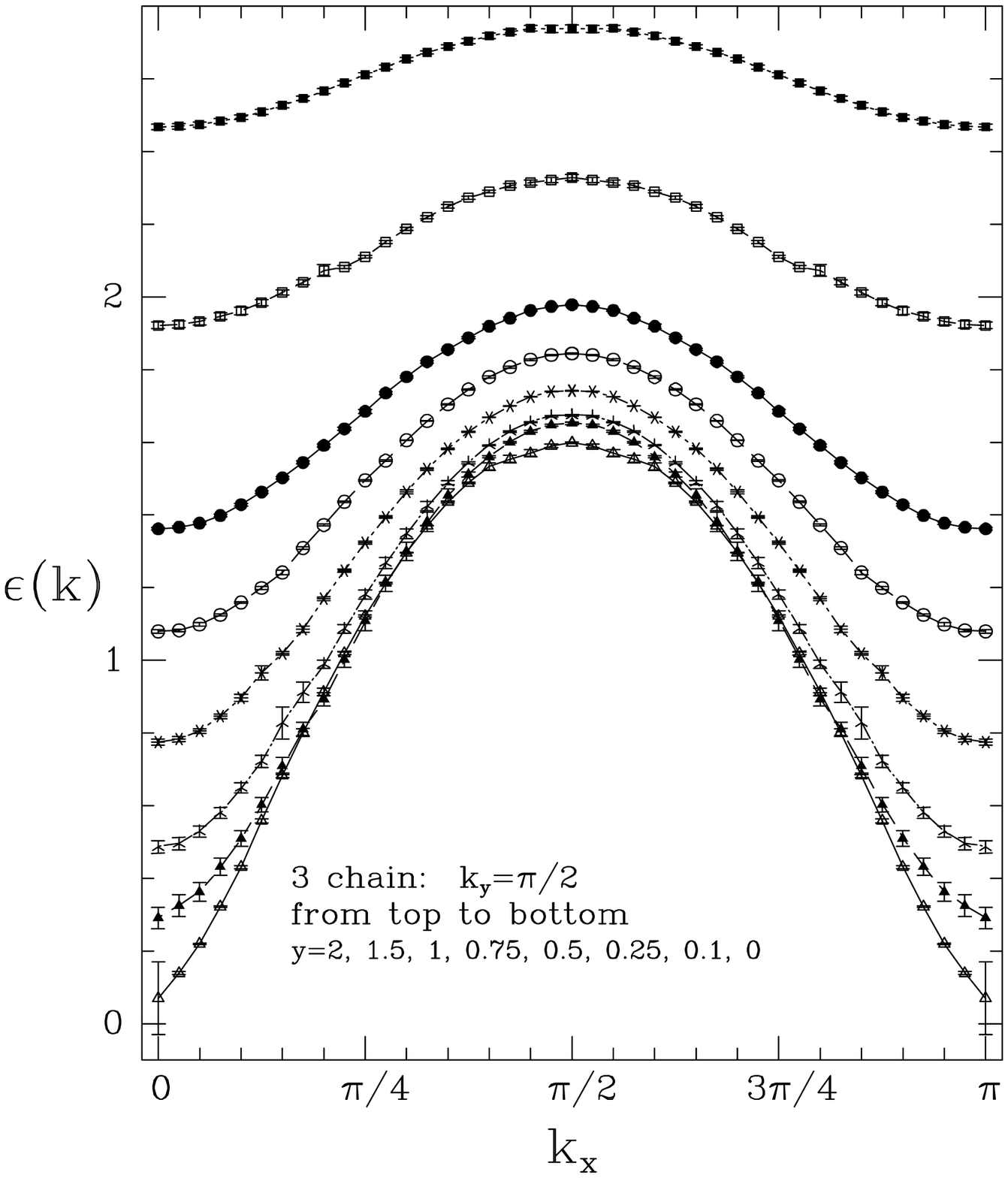,width=95mm}}
\caption{The dispersions of the spin-triplet excitated states of the 3-chain ladder
with antiferromagnetic interchain coupling $y=$2, 1.5, 1, 0.75, 0.5, 0.25, 0.1, 0,
 for $k_y=\pi/2$.
}
\label{fig:fig8}
\end{figure}

\begin{figure}[htb]
\vspace{9pt}
\makebox[80mm][l]{{\hskip 1.3pc} \psfig{file=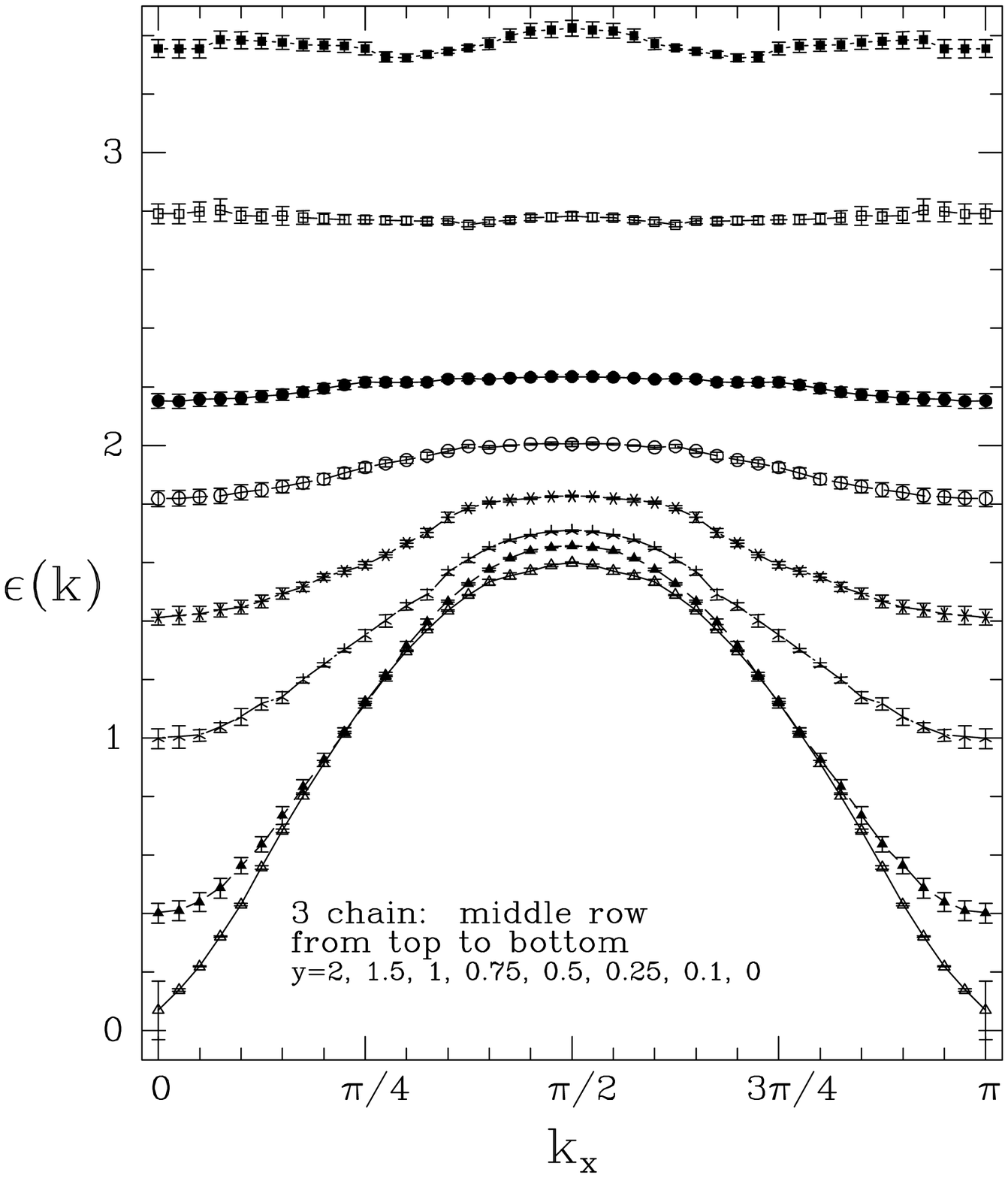,width=95mm}}
\caption{The dispersions of the spin-triplet excitated states of the 3-chain ladder
with antiferromagnetic interchain coupling $y=$2, 1.5, 1, 0.75, 0.5, 0.25, 0.1, 0,
 for the middle excitation band.
}
\label{fig:fig9}
\end{figure}

\begin{figure}[htb]
\vspace{9pt}
\makebox[80mm][l]{{\hskip 1.3pc} \psfig{file=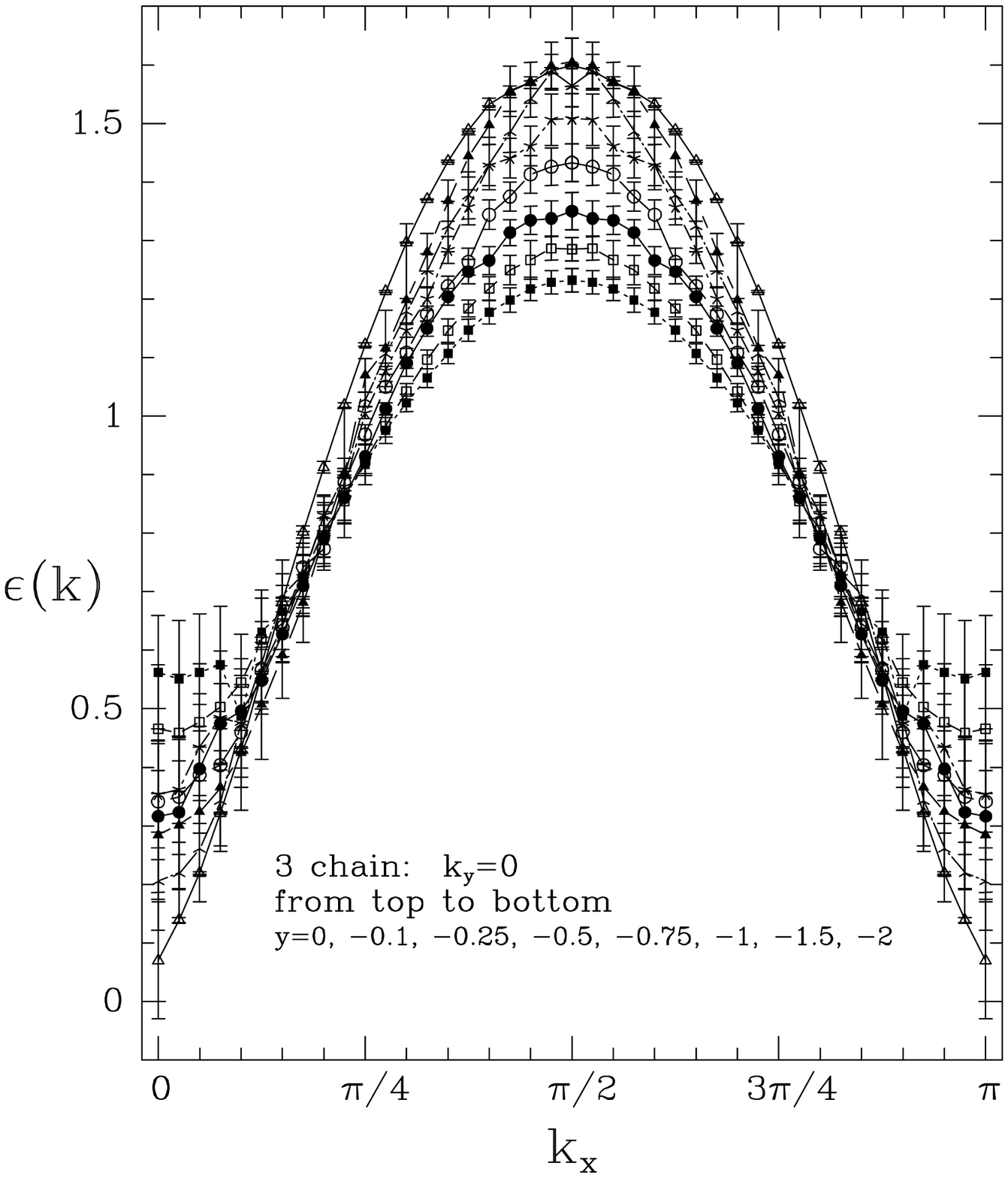,width=95mm}}
\caption{The dispersions of the spin-triplet excitated states of the 3-chain ladder 
with ferromagnetic interchain coupling $y=$0, -0.1, -0.25, -0.5, -0.75, -1, -1.5, -2
 for $k_y=0$.
}
\label{fig:fig10}
\end{figure}

\begin{figure}[htb]
\vspace{9pt}
\makebox[80mm][l]{{\hskip 1.3pc} \psfig{file=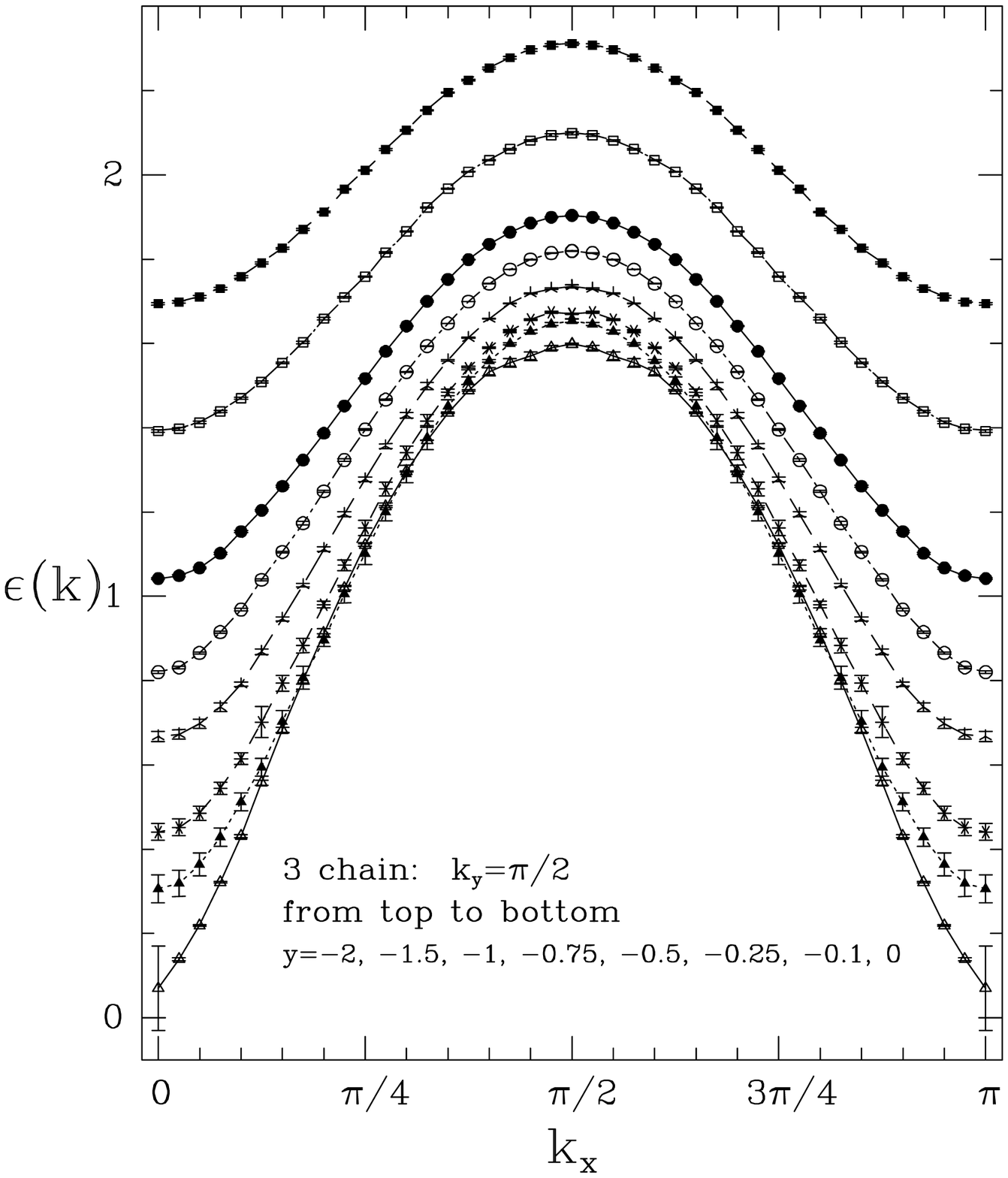,width=95mm}}
\caption{The dispersions of the spin-triplet excitated states of the 3-chain ladder
with ferromagnetic interchain coupling $y=$-2, -1.5, -1, -0.75, -0.5, -0.25, -0.1, 0,
 for $k_y=\pi/2$.
}
\label{fig:fig11}
\end{figure}

\begin{figure}[htb]
\vspace{9pt}
\makebox[80mm][l]{{\hskip 1.3pc} \psfig{file=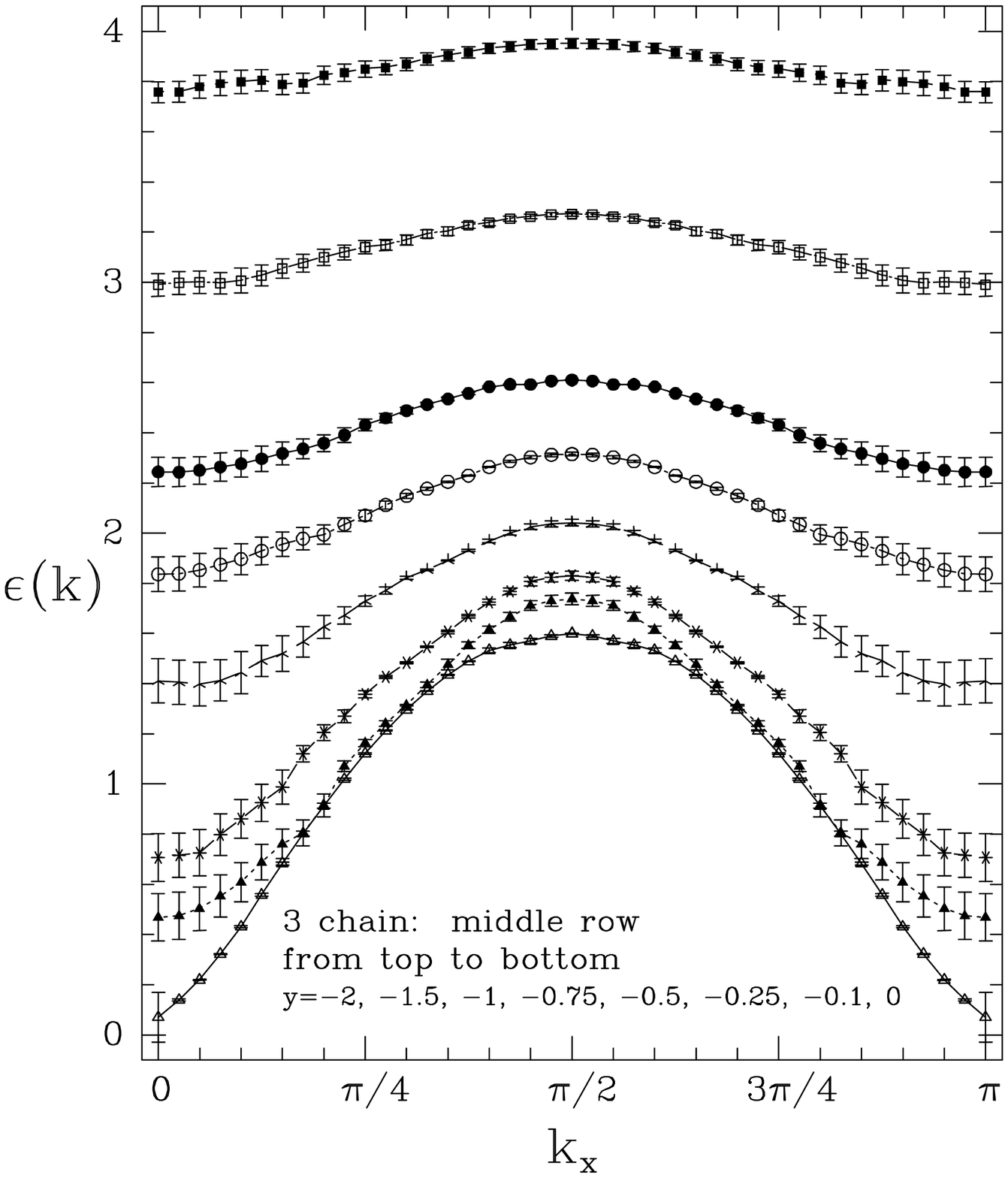,width=95mm}}
\caption{The dispersions of the spin-triplet excitated states of the 3-chain ladder
with ferromagnetic interchain coupling $y=$-2, -1.5, -1, -0.75, -0.5, -0.25, -0.1, 0,
 for the middle excitation band.
}
\label{fig:fig12}
\end{figure}
 
\begin{figure}[htb]
\vspace{9pt}
\makebox[80mm][l]{{\hskip 1.3pc} \psfig{file=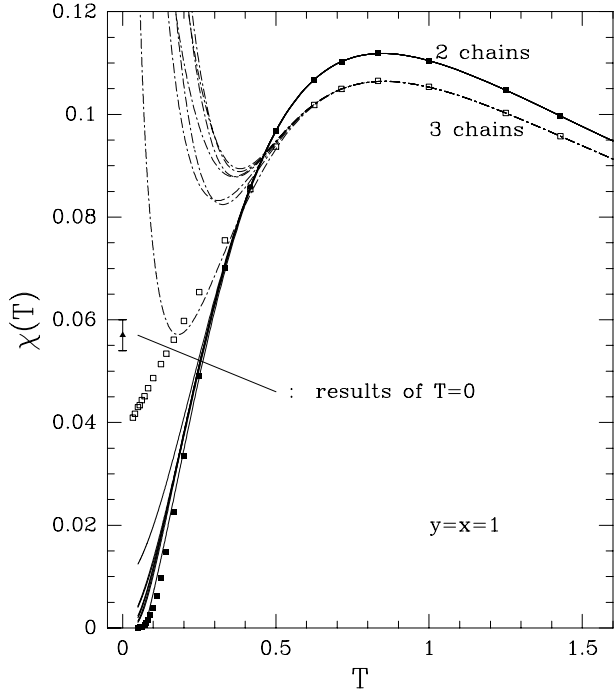,width=95mm}}
\caption{Susceptibility as a function of temperature for 2-chain and 3-chain
ladders from the high temperature series expansion, and the Ising expansion at $T=0$ (for 3-chain only),
 also shown are the QMC results of
Frischmuth\protect\cite{fri96} et al as the open symbols (for 2-chain) and filled symbols (for 3-chain)
for comparison,
}
\label{fig:fig13}
\end{figure}
 
\begin{figure}[htb]
\vspace{9pt}
\makebox[80mm][l]{{\hskip 1.3pc} \psfig{file=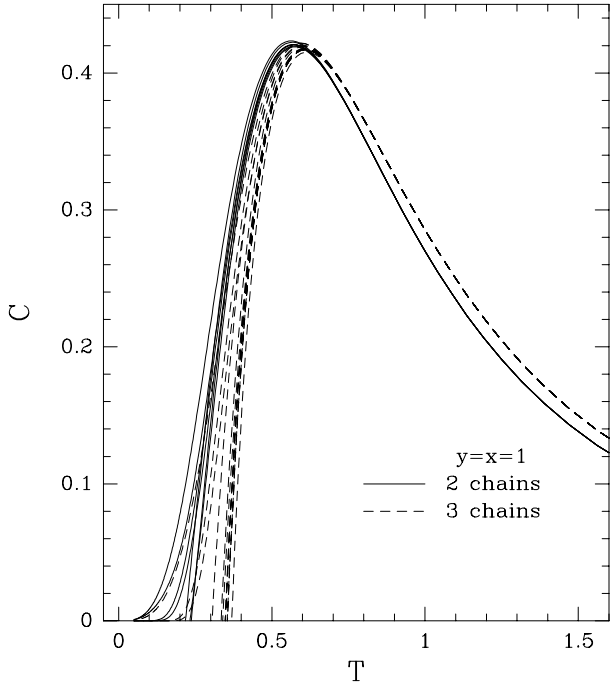,width=95mm}}
\caption{The specific heat as a function of temperature for 2-chain and 3-chain
ladders from the high temperature series expansion.
}
\label{fig:fig14}
\end{figure}
 
\widetext
\begin{table}
\squeezetable
\setdec 0.000000000000000
\caption{Series coefficients for the
ground-state energy per site $E_0/N$, the
staggered/colinear magnetization $M$, and staggered parallel
susceptibility $\chi_{/\!/}$. Coefficients of
$x^n$ are listed for both the
spin-${1\over 2}$ 2-chain and 3-chain ladders with $y=\pm 1$. } \label{tab1}
\begin{tabular}{rrrr}
 \multicolumn{1}{c}{n} &\multicolumn{1}{c}{$E_0/N$}
&\multicolumn{1}{c}{$M$} &\multicolumn{1}{c}{$\chi_{/\!/}$}  \\
\tableline
\\
\multicolumn{4}{l}{{\bf 2-chain ladders with $y=-1$}} \\
\multicolumn{1}{r}{0} &\multicolumn{1}{c}{$-$3/8}
&\multicolumn{1}{c}{1/2} &\multicolumn{1}{c}{0}  \\
  2 &\dec  $-$1.250000000000$\times 10^{-1}$ &\dec  $-$1.250000000000$\times 10^{-1}$ &\dec   2.500000000000$\times 10^{-1}$ \\
  4 &\dec  $-$1.562500000000$\times 10^{-2}$ &\dec  $-$6.944444444444$\times 10^{-2}$ &\dec   4.131944444444$\times 10^{-1}$ \\
  6 &\dec     2.443214699074$\times 10^{-3}$ &\dec     1.398473668981$\times 10^{-3}$ &\dec   9.964453526878$\times 10^{-2}$ \\
  8 &\dec  $-$3.391054004308$\times 10^{-3}$ &\dec  $-$4.293584111636$\times 10^{-2}$ &\dec   6.561704135203$\times 10^{-1}$ \\
 10 &\dec  $-$3.083929459431$\times 10^{-4}$ &\dec  $-$9.459983497527$\times 10^{-3}$ &\dec   2.592402875212$\times 10^{-1}$ \\
 12 &\dec  $-$5.758725813964$\times 10^{-4}$ &\dec  $-$2.517788286146$\times 10^{-2}$ &\dec   8.633367551243$\times 10^{-1}$ \\
 14 &\dec  $-$3.508609329659$\times 10^{-4}$ &\dec  $-$1.290819523741$\times 10^{-2}$ &\dec   5.194241039530$\times 10^{-1}$ \\
 16 &\dec  $-$3.658331257827$\times 10^{-4}$ &\dec  $-$2.220489673071$\times 10^{-2}$ &\dec   1.197591274920     \\
\\
\multicolumn{4}{l}{{\bf 2-chain ladders with $y=1$}} \\
\multicolumn{1}{r}{0} &\multicolumn{1}{c}{$-$3/8}
&\multicolumn{1}{c}{1/2} &\multicolumn{1}{c}{0}  \\
  2 &\dec  $-$1.875000000000$\times 10^{-1}$ &\dec  $-$1.875000000000$\times 10^{-1}$ &\dec   3.750000000000$\times 10^{-1}$ \\
  4 &\dec  $-$9.114583333333$\times 10^{-3}$ &\dec  $-$1.176215277778$\times 10^{-1}$ &\dec   9.751157407407$\times 10^{-1}$ \\
  6 &\dec  $-$6.564670138889$\times 10^{-3}$ &\dec  $-$1.298255449460$\times 10^{-1}$ &\dec   2.225074548290     \\
  8 &\dec  $-$1.011191788034$\times 10^{-2}$ &\dec  $-$2.567768064481$\times 10^{-1}$ &\dec   6.466470614293     \\
 10 &\dec  $-$4.438918908655$\times 10^{-3}$ &\dec  $-$3.308567310261$\times 10^{-1}$ &\dec   1.436049594755$\times 10^{1}$ \\
 12 &\dec  $-$1.149045736404$\times 10^{-2}$ &\dec  $-$6.754838729347$\times 10^{-1}$ &\dec   3.661634831290$\times 10^{1}$ \\
 14 &\dec  $-$9.014538657677$\times 10^{-3}$ &\dec  $-$1.059678748481     &\dec   8.236161272382$\times 10^{1}$ \\
 16 &\dec  $-$1.860424772341$\times 10^{-2}$ &\dec  $-$2.087462804052     &\dec   1.962459013257$\times 10^{2}$ \\
\\
\multicolumn{4}{l}{{\bf 3-chain ladders with $y=-1$}} \\
\multicolumn{1}{r}{0} &\multicolumn{1}{c}{$-$5/12}
&\multicolumn{1}{c}{1/2} &\multicolumn{1}{c}{0}  \\
  2 &\dec  $-$1.111111111111$\times 10^{-1}$ &\dec  $-$1.018518518519$\times 10^{-1}$ &\dec   1.913580246914$\times 10^{-1}$ \\
  4 &\dec  $-$9.126984126984$\times 10^{-3}$ &\dec  $-$2.765516082976$\times 10^{-2}$ &\dec   1.235451495119$\times 10^{-1}$ \\
  6 &\dec  $-$3.430007743753$\times 10^{-3}$ &\dec  $-$2.229002861112$\times 10^{-2}$ &\dec   1.702591936321$\times 10^{-1}$ \\
  8 &\dec  $-$4.479840217373$\times 10^{-4}$ &\dec  $-$6.778016397592$\times 10^{-3}$ &\dec   9.228960192872$\times 10^{-2}$ \\
 10 &\dec  $-$1.150888174205$\times 10^{-3}$ &\dec  $-$1.348552579039$\times 10^{-2}$ &\dec   1.814452046940$\times 10^{-1}$ \\
 12 &\dec  $-$5.493987828590$\times 10^{-4}$ &\dec  $-$9.549725222639$\times 10^{-3}$ &\dec   1.785680892107$\times 10^{-1}$ \\
\\
\multicolumn{4}{l}{{\bf 3-chain ladders with $y=1$}} \\
\multicolumn{1}{r}{0} &\multicolumn{1}{c}{$-$5/12}
&\multicolumn{1}{c}{1/2} &\multicolumn{1}{c}{0}  \\
  2 &\dec  $-$1.777777777778$\times 10^{-1}$ &\dec  $-$1.551851851852$\times 10^{-1}$ &\dec   2.766913580247$\times 10^{-1}$ \\
  4 &\dec     2.099353321576$\times 10^{-3}$ &\dec  $-$3.010635013951$\times 10^{-2}$ &\dec   2.520792079948$\times 10^{-1}$ \\
  6 &\dec  $-$5.486291040711$\times 10^{-3}$ &\dec  $-$4.567138485811$\times 10^{-2}$ &\dec   4.781602664625$\times 10^{-1}$ \\
  8 &\dec  $-$9.770232257758$\times 10^{-4}$ &\dec  $-$2.676534614009$\times 10^{-2}$ &\dec   5.289125611753$\times 10^{-1}$ \\
 10 &\dec  $-$8.540222317364$\times 10^{-4}$ &\dec  $-$2.413110830174$\times 10^{-2}$ &\dec   6.371032432384$\times 10^{-1}$ \\
 12 &\dec  $-$7.963887426932$\times 10^{-4}$ &\dec  $-$3.026105625065$\times 10^{-2}$ &\dec   1.024475965994     \\
\end{tabular}
\end{table}

\setdec 0.00000000000
\begin{table}
\squeezetable
\caption{Series coefficients for the perpendicular susceptibility
$\chi_\perp$. Coefficients of $x^n$ are listed for 2-chain and 3-chain ladders with $y=\pm 1$.}\label{tab2}
\begin{tabular}{rrrrr}
\multicolumn{1}{r}{n} &\multicolumn{1}{c}{2-chain $y=-1$}
&\multicolumn{1}{c}{2-chain $y=1$} &\multicolumn{1}{c}{3-chain $y=-1$}
&\multicolumn{1}{c}{3-chain $y=1$} \\
\tableline
\multicolumn{1}{r}{0} & \multicolumn{1}{c}{1/3} & \multicolumn{1}{c}{1/3} &
\multicolumn{1}{c}{11/36} & \multicolumn{1}{c}{11/36}    \\
\multicolumn{1}{r}{1} & \multicolumn{1}{c}{$-$2/9} & \multicolumn{1}{c}{$-$1/2}
& \multicolumn{1}{c}{$-$1/6} & \multicolumn{1}{c}{$-$13/30}    \\
  2 &\dec   1.481481481481$\times 10^{-2}$ &\dec   5.666666666667$\times 10^{-1}$ &\dec   1.266534391534$\times 10^{-2}$ &\dec   4.652226631393$\times 10^{-1}$ \\
  3 &\dec   5.308641975309$\times 10^{-3}$ &\dec  $-$7.143518518519$\times 10^{-1}$ &\dec  $-$1.112134668682$\times 10^{-2}$ &\dec  $-$5.160474965706$\times 10^{-1}$ \\
  4 &\dec  $-$2.120002939447$\times 10^{-2}$ &\dec   7.544920634921$\times 10^{-1}$ &\dec   1.134331580417$\times 10^{-2}$ &\dec   5.434191826896$\times 10^{-1}$ \\
  5 &\dec   7.283052502380$\times 10^{-2}$ &\dec  $-$9.203876999874$\times 10^{-1}$ &\dec  $-$2.787971676358$\times 10^{-2}$ &\dec  $-$6.037870527565$\times 10^{-1}$ \\
  6 &\dec  $-$6.507519406167$\times 10^{-2}$ &\dec   9.542443240384$\times 10^{-1}$ &\dec   3.588831815932$\times 10^{-2}$ &\dec   6.192643274088$\times 10^{-1}$ \\
  7 &\dec  $-$9.850515124711$\times 10^{-3}$ &\dec  $-$1.220670122297     &\dec  $-$3.496516710785$\times 10^{-2}$ &\dec  $-$6.538548959720$\times 10^{-1}$ \\
  8 &\dec  $-$3.321642365298$\times 10^{-3}$ &\dec   1.243330662897     &\dec   3.361956127668$\times 10^{-2}$ &\dec   6.836898452099$\times 10^{-1}$ \\
  9 &\dec   3.225303151489$\times 10^{-2}$ &\dec  $-$1.570424359869     &\dec  $-$4.583945317287$\times 10^{-2}$ &\dec  $-$7.207963780768$\times 10^{-1}$ \\
 10 &\dec  $-$2.164319811625$\times 10^{-2}$ &\dec   1.612928972471     &\dec   4.084885561076$\times 10^{-2}$ &\dec   7.446415246323$\times 10^{-1}$ \\
 11 &\dec  $-$2.229164879943$\times 10^{-2}$ &\dec  $-$2.181953857121     &\dec  $-$4.495109043570$\times 10^{-2}$ &\dec  $-$7.944909220697$\times 10^{-1}$ \\
 12 &\dec  $-$8.386607266663$\times 10^{-3}$ &\dec   2.245984866808         \\
 13 &\dec   1.980513341011$\times 10^{-2}$ &\dec  $-$3.034629087270         \\
 14 &\dec  $-$1.128439840082$\times 10^{-2}$ &\dec   3.158070079722         \\
 15 &\dec  $-$1.625818723347$\times 10^{-2}$ &\dec  $-$4.546560413718       \\
\end{tabular}
\end{table}

\setdec 0.00000000000
\begin{table}
\squeezetable
\caption{Series coefficients for the dimer expansion of the 2-chain triplet 
spin-wave excitation spectrum $\epsilon (k_x, k_y=\pi) = y \sum_{n,m} a_{n,m} (1/y)^n \cos (m k_x)$. Nonzero coefficients $a_{n,m}$ up to order $n=8$ are listed.}\label{tab3}
\begin{tabular}{rr|rr|rr|rr}
\multicolumn{1}{c}{(n,m)} &\multicolumn{1}{c|}{$a_{n,m}$}
&\multicolumn{1}{c}{(n,m)} &\multicolumn{1}{c|}{$a_{n,m}$}
&\multicolumn{1}{c}{(n,m)} &\multicolumn{1}{c|}{$a_{n,m}$}
&\multicolumn{1}{c}{(n,m)} &\multicolumn{1}{c}{$a_{n,m}$} \\
 ( 0, 0) &\dec   1.000000000     &( 5, 1) &\dec $-$2.031250000$\times 10^{-1}$ &( 3, 3) &\dec   1.250000000$\times 10^{-1}$ &( 5, 5) &\dec   5.468750000$\times 10^{-2}$ \\
 ( 2, 0) &\dec   7.500000000$\times 10^{-1}$ &( 6, 1) &\dec   9.375000000$\times 10^{-2}$ &( 4, 3) &\dec   1.250000000$\times 10^{-1}$ &( 6, 5) &\dec   7.812500000$\times 10^{-2}$ \\
 ( 3, 0) &\dec   3.750000000$\times 10^{-1}$ &( 7, 1) &\dec   3.293457031$\times 10^{-1}$ &( 5, 3) &\dec $-$9.375000000$\times 10^{-2}$ &( 7, 5) &\dec $-$6.042480469$\times 10^{-2}$ \\
 ( 4, 0) &\dec $-$2.031250000$\times 10^{-1}$ &( 8, 1) &\dec   2.555847168$\times 10^{-1}$ &( 6, 3) &\dec $-$3.164062500$\times 10^{-1}$ &( 8, 5) &\dec $-$2.657165527$\times 10^{-1}$ \\
 ( 5, 0) &\dec $-$6.250000000$\times 10^{-1}$ &( 2, 2) &\dec $-$2.500000000$\times 10^{-1}$ &( 7, 3) &\dec $-$2.222900391$\times 10^{-1}$ &( 6, 6) &\dec $-$4.101562500$\times 10^{-2}$ \\
 ( 6, 0) &\dec $-$5.000000000$\times 10^{-1}$ &( 3, 2) &\dec $-$2.500000000$\times 10^{-1}$ &( 8, 3) &\dec   2.752685547$\times 10^{-1}$ &( 7, 6) &\dec $-$6.835937500$\times 10^{-2}$ \\
 ( 7, 0) &\dec   2.966308594$\times 10^{-1}$ &( 4, 2) &\dec $-$3.125000000$\times 10^{-2}$ &( 4, 4) &\dec $-$7.812500000$\times 10^{-2}$ &( 8, 6) &\dec   4.957580566$\times 10^{-2}$ \\
 ( 8, 0) &\dec   1.120300293     &( 5, 2) &\dec   2.031250000$\times 10^{-1}$ &( 5, 4) &\dec $-$9.375000000$\times 10^{-2}$ &( 7, 7) &\dec   3.222656250$\times 10^{-2}$ \\
 ( 1, 1) &\dec   1.000000000     &( 6, 2) &\dec   1.718750000$\times 10^{-1}$ &( 6, 4) &\dec   7.128906250$\times 10^{-2}$ &( 8, 7) &\dec   6.152343750$\times 10^{-2}$ \\
 ( 3, 1) &\dec $-$2.500000000$\times 10^{-1}$ &( 7, 2) &\dec $-$1.728515625$\times 10^{-1}$ &( 7, 4) &\dec   2.690429688$\times 10^{-1}$ &( 8, 8) &\dec $-$2.618408203$\times 10^{-2}$ \\
 ( 4, 1) &\dec $-$3.125000000$\times 10^{-1}$ &( 8, 2) &\dec $-$5.047454834$\times 10^{-1}$ &( 8, 4) &\dec   1.690521240$\times 10^{-1}$ &  &    \\
\end{tabular}
\end{table}

\setdec 0.00000000000
\begin{table}
\squeezetable
\caption{Series coefficients for high temperature series
expansion of the uniform susceptibility
$\chi (T) = \beta \sum_i c_i \beta^i/(n_l 2^{i+4} i!)$,
and the specific heat $C (T) = \beta^2 \sum_i c_i \beta^i/(n_l 2^{i+5} i!)$.
 Coefficients $c_i$ are listed for 2-chain and 3-chain ladders with $y=1$.}\label{tab4}
\begin{tabular}{rrrrr}
\multicolumn{1}{r}{i} &\multicolumn{1}{c}{$\chi (T)$ for 2-chain}
&\multicolumn{1}{c}{$\chi (T)$ for 3-chain} 
 &\multicolumn{1}{c}{$C (T)$ for 2-chain}
&\multicolumn{1}{c}{$C (T)$ for 3-chain}  \\
\tableline
     0 & 8   &  12   & 36   &  60 \\
     1 &  $-$12  & $-$20 & 72   &  120 \\
     2 & 12   & 28 &  $-$270   &  $-$522  \\
     3 & 6   & $-$20 & $-$2640   &  $-$5040 \\
     4 &  $-$20  &  4 &  90   &  3270 \\
     5 &  $-$162  & $-$160 &  141876   &  318780\\
     6 &  $-$630  & $-$1052 &  580797   &  1075767\\
     7 &  9991  &  17298 &  $-$10663200   &  $-$28792032 \\
     8 & 88228  &  80468 &  $-$118074186  &  $-$291518730 \\
     9 &  $-$779322  & $-$1467200 &  946669020   &  3061122900\\
    10 &  $-$13957358  & $-$12792822 &  26078160405  &  76820424879 \\
    11 & 55717397   & 165603440 &  $-$42521155560    & $-$195632449272\\
    12 & 2827957594  &  2955180058  & $-$6789937647207   &  $-$22502126499801\\
    13 & 4867299659   &$-$24526691326    \\
    14 &  $-$687967034169  & $-$924449102836    \\
\end{tabular}
\end{table}

\end{document}